\documentclass[%
 reprint,
 amsmath,amssymb,
 aps,
]{revtex4-2}

\usepackage{graphicx}%
\usepackage{dcolumn}%
\usepackage{bm}%

\begin{document}

\title{Single-step transmission matrix retrieval\\ for fast imaging through multi-mode fibers} %

\author{Daniele Ancora$^{1}$}
\altaffiliation[correspondence to: daniele.ancora@uniroma1.it ]{\email{daniele.ancora@uniroma1.it}}%
\author{Lorenzo Dominici$^2$}%
\author{Antonio Gianfrate$^2$}%
\author{Paolo Cazzato$^2$}%
\author{Milena De Giorgi$^2$}%
\author{Dario Ballarini$^2$}%
\author{Daniele Sanvitto$^2$}%
\author{Luca Leuzzi$^{1,3}$}

\affiliation{%
$^1$Department of Physics, Università di Roma la Sapienza, Piazzale Aldo Moro 5, I-00185 Rome, Italy}%

\affiliation{$^2$Institute of Nanotechnology, Consiglio Nazionale delle Ricerche (CNR-NANOTEC), Via Monteroni, I-73100 Lecce, Italy}%

\affiliation{$^3$Institute of Nanotechnology, Soft and Living Matter Laboratory, Consiglio Nazionale delle Ricerche (CNR-NANOTEC), Piazzale Aldo Moro 5, I-00185 Rome, Italy}%

\date{\today}%

\begin{abstract}
Recovering the transmission matrix of a disordered medium is a challenging problem in disordered photonics.
Usually, its reconstruction relies on a complex inversion that aims at connecting a fully-controlled input to the deterministic interference of the light field scrambled by the device.
At the moment, iterative phase-retrieval protocols provide the fastest reconstructing frameworks, converging in a few tens of iterations.
Exploiting the knowledge of speckle correlations, we construct a new phase retrieval algorithm that reduces the computational cost to a single iteration.
Besides being faster, our method is effective also using less measurements than state-of-the-art protocols.
Thanks to reducing computation time by one order of magnitude, our result can be groundbreaking for real-time optical operations in medical imaging.

\end{abstract}

\maketitle

\section{\label{sec:Intro} Introduction}
One of the greatest efforts in modern optics is to exploit disordered structures for imaging and focusing through (or, perhaps, inside) optical materials.
The revolution began with the usage of light shaping devices to manipulate the light field, observing how the turbid medium reacts to controlled excitations \cite{vellekoop2007focusing}.
Initially, imaging procedures were mainly accomplished by taking advantage of the memory effect \cite{bertolotti2012non} whereas, for focusing, feedback-based genetic algorithms were common \cite{vellekoop2010exploiting, Conkey:12}.
However, the light propagation through a scattering device can be modeled as a linear process, in which a -complicated and unknown- transmission matrix (TM) deterministically controls how the light is transported by the medium \cite{van2010information, popoff2010measuring}.
Measuring the TM became rapidly an effective approach in particular for the exploitation of turbid devices as standard optical tools, stimulating intense research activities \cite{RevModPhys.89.015005}.
Because of this, researchers developed many approaches for the recovery of the TM in different scenarios.
Among others, holography requires to have access to both edges via an interferometric configuration \cite{ploschner2015seeing}.
Although complex to implement, these setups provided thriving results.
For holographic imaging, the usage of the reference arm permitted to obtain complete control over the image transmission through disordered channels \cite{goorden2014superpixel}.
Recently, a compressed sensing approach was used to recover the optical TM of multi-mode fiber with a reduced number of probe measurements \cite{li2021compressively}.
Also, the memory effect was exploited to assist the recovery of the TM in a multi-mode fiber \cite{li2021memory}.

\begin{figure}[t!]
\includegraphics[trim=0 100 0 0, clip, width=1\linewidth]{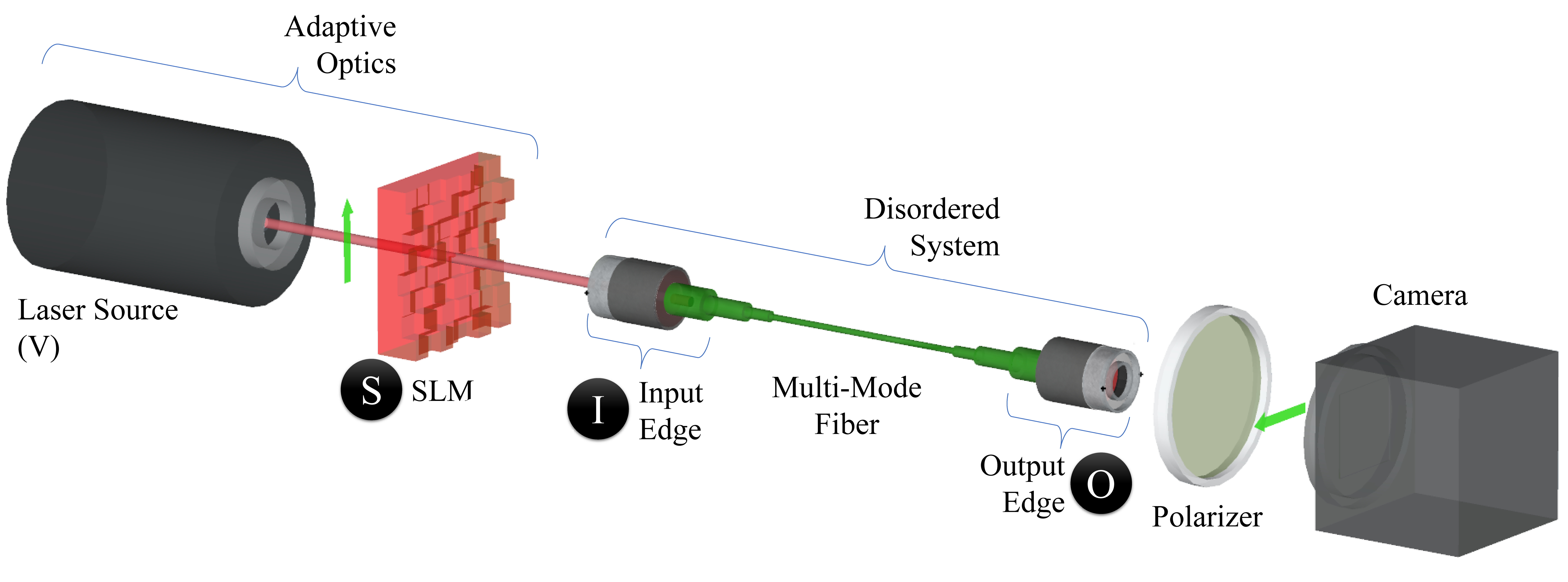}%
\caption{\label{fig:setup} Scheme of the setup used for our imaging experiments.
A laser source (vertical polarization) is modulated into a probing pattern using a spatial light modulator (SLM).
Once modulated, the field is projected onto the input edge of a multi-mode fiber (I).
The light which trespasses the disordered medium is imaged at the output edge (O) by a standard camera sensor (horizontal polarization).}
\end{figure}

Although the holographic approach allows accurate characterization of the transmission, it is of difficult applicability in the real-measurement scenario.
The presence of the reference arm, external to the actual fiber bundle, hinders the miniaturization of the optical device.
These results, however, drive the investigation in non-interferometric configurations.
As schematized in Fig. \ref{fig:setup}, the simple imaging setup consists in sending a given pattern with a light shaping device in the input side of a disordered medium (here, we are interested in multi-mode fibers) and recording its transmission at the output edge.
In the attempt to simplify the recovery in such a simple configuration, a portion of the SLM was kept fix to be used as a self-reference for the characterization \cite{popoff2010image, popoff2010measuring}.
Random measurements also provided excellent frameworks for the TM reconstructions with the reference-less prVBEM Bayesian approach \cite{dremeau2015reference}, or with Gerchberg-Saxton iterative approaches \cite{huang2021generalizing}.
Furthermore, measurements using Hadamard basis have permitted approximate the TM with real-valued entries \cite{zhao2020seeing}.
Currently, all the methods are supervised inversions in the sense that a known pattern is sent and the related output field is measured.
By repeating this procedure several times, one can gain complete knowledge over the scattering process that the light field undergoes during propagation.
Usually, the higher the number of patterns that are sampled the better the estimation of the matrix.
Once recovered, the TM can be used to produce a focus in a user-specified position, or to invert the scattering process for the recovery of the images.

In both applications, the statistics of the speckle pattern produced by the transmission of light through the random medium tells us information about its optical response \cite{devaud2020speckle}.
A fundamental property is that a point source produces a speckle whose auto-correlation sets the optimal resolution achievable by the system \cite{di2016tailored, di2016tailoring}.
This is connected to the point spreading function in the propagation through homogeneous media.
When analyzing a speckled image, there are regions of coherence generally larger than a single pixel.
To date, every method for reconstructing the TM assumes a "single-pixel camera" approach, which delineates a separable problem at the output plane.
Each calibration pattern sent at the input contributes to the output of a single-pixel independently.
This simplistic assumption uses less memory and is very useful for the algorithm parallelization, but it discards useful neighboring interactions.
In our study, we exploit the physical information provided by the speckle statistics in order to aid the reconstruction of the TM.
We implement it by proposing a modified, non-interferometric, Gerchberg-Saxton (GS) phase retrieval (PR) approach for imaging through disorder.
GS implementations have, in fact, the advantage of being the fastest framework for the reconstruction of the transmission matrix \cite{rodriguez2013oversampling}.
In a nutshell, our modification consists of linking adjacent pixels in the output plane by adding a step of image convolution with a tunable kernel.
Our idea takes inspiration from the oversampling smoothness protocol that was proposed to regularize the solution of the phase retrieval problem in Fourier imaging \cite{rodriguez2013oversampling}.
Here, instead, the smoothing step is used to couple adjacent output pixels according to the physical connection described by the average speckle size.
To test the effectiveness of our modification, we decided to tackle the problem of imaging through multi-mode fiber, performing a complete study on the number of patterns and iterations needed to achieve optimal reconstruction results.
Compared to state-of-the-art methods, our algorithm converges within a single iteration and can reconstruct images by using undersampled measurements.
We begin our study by describing the protocol and the non-interferometric experimental setup that we used for imaging.
After this, we will present a numerical study to characterize the behavior of our method, discussing the results and further perspectives.

\section{Materials and Methods}\label{sec:MatMet}
As described by Goodman \cite{goodman2007speckle}, the speckles distributed in a pattern produced by a scattering medium have an average size estimated via its auto-correlation.
For fully developed speckles, it turns to be a sharply peaked function, essentially being the auto-correlation of a broadband noise \cite{katz2014non}.
If we call $S$ the camera-recorded speckle pattern, this translates to the autocorrelation in the plane $S \star S \approx \delta$, the Dirac $\delta$-distribution.
Here, we denote with $\star$ the cross-correlation operator.
Within the memory effect range approach \cite{katz2014non, bertolotti2012non}, this property was exploited to perform hidden imaging based on direct speckle observation.
Experimentally, however, we are far away from obtaining a delta function.
Not having a point-like auto-correlation implies that neighboring modes detected by the camera are not independent completely but, instead, are spatially correlated on the plane.
Specifically, we can assume that the speckles exhibit a Gaussian auto-correlation, $S \star S = g\left( \Sigma \right)$, which can be fitted by recovering its standard deviation $\Sigma$.
In the following, we exploit this information to allow our method to converge faster than state-of-art implementations.
So far, any approach proposed to retrieve the TM considers each pixel in the output edge independent of each other.
Here, instead, we take into account neighboring interaction by introducing a step in the GS algorithm that couples the output pixels via the expected point-spread function (PSF) resolution of the system, providing imaging benefits in terms of reconstruction efficiency.

\subsection{\label{ssec:PhaseRet} Phase retrieval description}
\begin{figure}[b!]
\includegraphics[trim=0 0 0 0, clip, width=1.\linewidth]{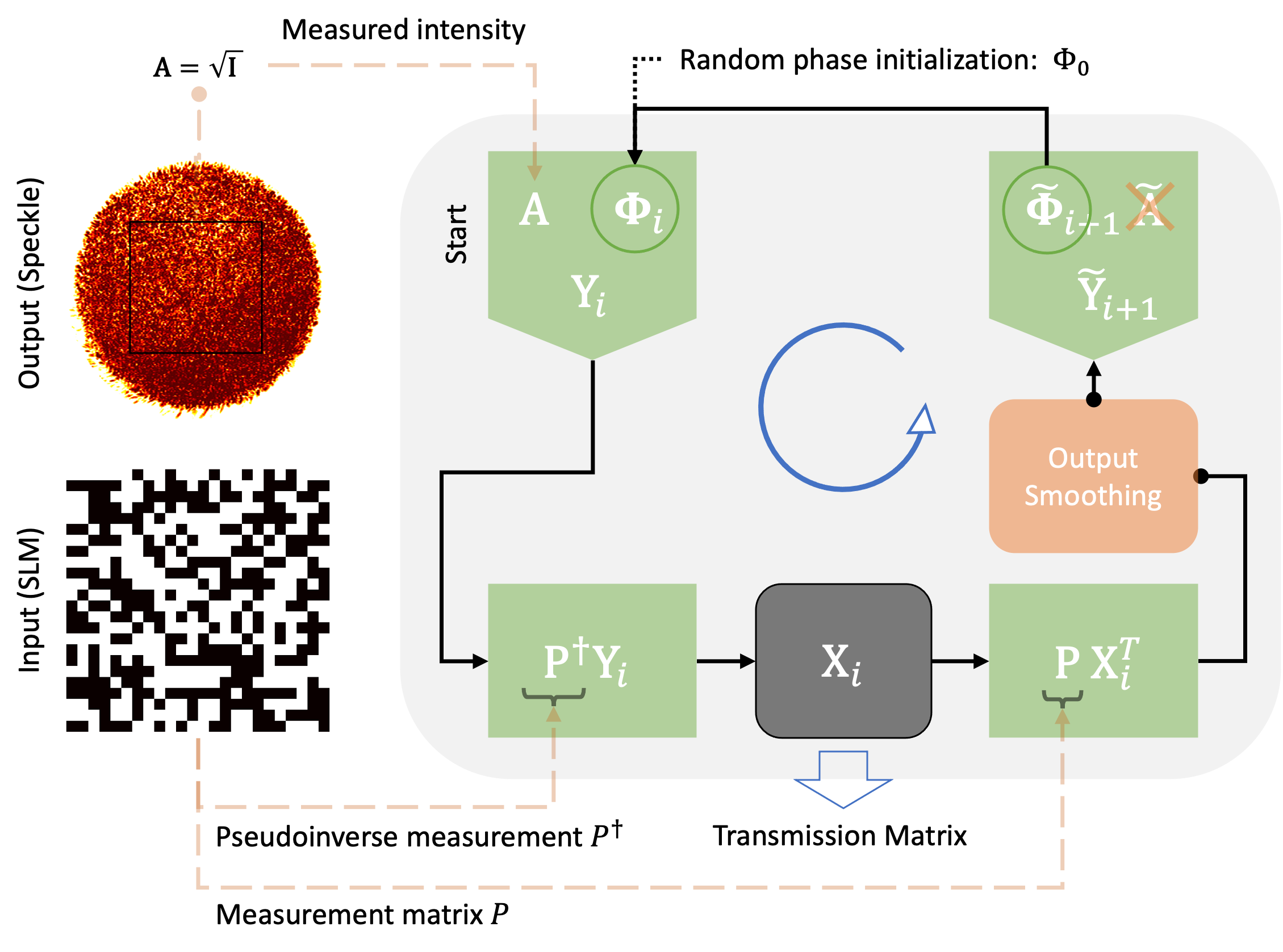}
\caption{\label{fig:schemeGS}
Scheme of the Gerchberg-Saxton phase retrieval protocols used in our manuscript.
The orange box indicates where the smoothing operation takes place.
Without this operation, the phase retrieval protocol is the same described in \cite{huang2021generalizing}.}
\end{figure}

Phase retrieval is a class of algorithms aiming at the recovery of the phase of a wavefront given a set of intensity measurements.
The problem that we aim at describing can be formulated as a linear field combination that results in the formation of a disordered speckle pattern on the output edge of the multi-mode (MM) fiber.
We describe the problem as follows.
We send a set of random binary patterns of size $N_i = L \times L$ from the input edge modulated by the SLM.
Even if the patterns are bi-dimensional, we store them as single dimension arrays within the probing matrix $\textbf{P}$, so that
each row in $\textbf{P}$ represents a given pattern.
In general, we consider a variable number of measurements $M$ (each corresponding to a different input pattern) so that $\textbf{P}$ has a dimension of $M\times N_i$.
We assume that the disordered medium can be described by an unknown complex transmission matrix $\textbf{X}$, which scrambles the input $\textbf{P}$ into the output measurements $\textbf{Y}$, each measurement displaying $N_o$ output pixels.
The underlined linear problem can be written as:
\begin{equation}
    \textbf{Y} = \textbf{P} \textbf{X}^T.
\end{equation}
The matrix $\textbf{X}^T$ has a size of $N_i \times N_o$ and the output matrix $\textbf{Y}$ has dimensions of $M \times N_o$.
The goal of a phase retrieval problem is to find $\textbf{X}$, given the probing matrix $\textbf{P}$ and the modulus of the measurements $\left| \textbf{Y} \right|$.
In our case, in fact, the camera records the field intensity $\textbf{I} = \left| \textbf{Y} \right|^2$, which stores every speckle pattern $S$ obtained in response to each input pattern.
We refer to it as phase retrieval because the method retrieves the phase of the output $\textbf{Y}$ and, consequently, lets us estimate $\textbf{X}$.
As already mentioned, this kind of problem was tackled in several ways, such as with the
EKF-MSSM approach \cite{huang2020retrieving},
SDP algorithm  \cite{n2017controlling, n2018mode},
prVBEM \cite{dremeau2015reference, zhao2018bayes, deng2018characterization},
GAMP \cite{schniter2014compressive} and its extension GVAMP \cite{sharma2019inverse}.
However, these classes of algorithms are computationally demanding, whereas GS approaches represent the fastest and workflow efficient alternatives, as thoroughly discussed by Huang et al. \cite{huang2021generalizing}.
Yet, the GS phase retrieval is a simple and efficient protocol based on an iterative approach, which we schematically describe in Fig. \ref{fig:schemeGS}.
The method begins by assigning a random phase ${{\mathbf{\Phi}}_0}$ to the measured intensity  to form the (complex) estimated output observation, $\textbf{Y}_0 = \textbf{A} e^{{\mathbf{\Phi}}_0}$.
Here, $\textbf{A}=\sqrt{\textbf{I}}$ is the squared root of the intensity detected on the camera.
Since the method is based on forward and backward application of the measurement patterns, we pre-compute the Moore-Penrose pseudo-inverse \cite{benisrael2003generalized} of the matrix $\mathbf{P}$, denoting it with $\mathbf{P}^\dagger$.

Compared to standard GS approaches, here, we describe the modification introduced by our phase retrieval protocol.
The method that we propose consists of 5 steps schematized in Fig. \ref{fig:schemeGS}:
\begin{enumerate}
    \item We form an estimation of the complex output $\mathbf{Y}_i$ by keeping the recorded modulus $\textbf{A}$ and associating a phase from the previous iteration. The initial phase $\mathbf{\Phi}_0$ is chosen from a uniform random distribution.
    \item We compute the new guess for the transmission matrix as $\mathbf{X}_i = \mathbf{P}^\dagger \mathbf{Y}_i$.
    \item We let the sequence of input patterns $\mathbf{P}$ to propagate through the retrieved TM, and we calculate the new output estimation as $\widetilde{\mathbf{Y}}_i = \mathbf{P} \mathbf{X}_i^T$.
    \item We carry out a convolution of this $\widetilde{\mathbf{Y}}_i$ with a Gaussian kernel setting its variance as a function of the iteration number.
    \item We keep the phase of the estimated output, as $\widetilde{\mathbf{\Phi}}_{i+1} = \text{Arg}\{ {\widetilde{\mathbf{Y}}_{i+1}} \}$.
    We pass this phase to step 1. of the next iteration.
\end{enumerate}
We refer to our method as SmoothGS (where appropriate, also abbreviated to SGS).
Without step 4, the iteration described is a standard GS phase retrieval \cite{huang2021generalizing}.
As a rule of thumb, we decide to vary the size of the neighboring interaction by controlling the sigma of the Gaussian kernel.
The first iteration starts with half the standard deviation determined by the fit of the auto-correlation with a Gaussian function, $\sigma=\Sigma/2$ (see App. \ref{app:C}).
Such kernel $g\left( \sigma \right)$ represents a Gaussian ensemble-average of the speckle size observed in the fiber.
The value is linearly decreased to reach $\sigma = 0.1 px$ that approximates the Gaussian as a delta function.
In a discretized kernel, the latter corresponds to an image with only the central pixel having a non-null value, whereas all the surrounding is set to zero.
In this way, we allow for a strong neighboring coupling at the beginning of the iteration, and we weaken its effect as the iteration proceeds.
After a given number of iteration steps, we are left with the TM that describes the system.

On the other hand, the imaging procedure consists in sending an unknown pattern to the input edge and reconstructing it; based on the speckle recorded at the output and the estimated TM.
The problem to be solved, in this case, is analogous to the previous one:
\begin{equation}
    \left(\textbf{Y}'\right)^T = \textbf{X} \left(\textbf{P}'\right)^T,
\end{equation}
where $\left(\textbf{Y}'\right)$ is the set of the observed speckle patterns and $\left(\textbf{P}'\right)$ are the unknown images on the input edge which generated it.
For our study, we implement the double phase retrieval method \cite{dremeau2015reference, rajaei2016intensity}.
The first problem retrieves the transmission matrix; the second carries out the image reconstruction.

\subsection{\label{ssec:Setup} Experimental measurements}
\begin{figure}[t!]
\includegraphics[width=1.\linewidth]{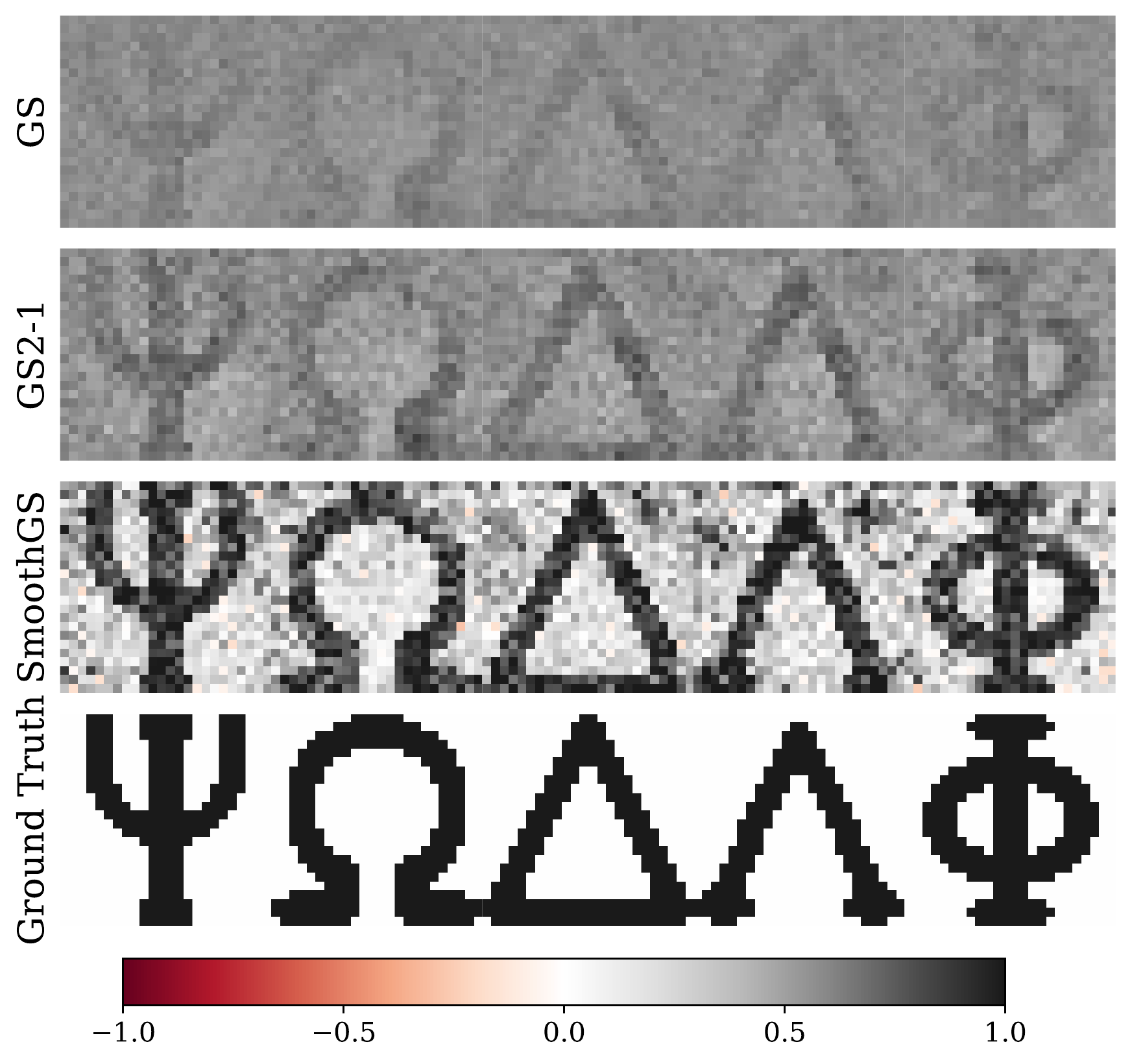}
\caption{\label{fig:symbolsRec} Imaging results for five input symbols obtained after a single iteration with $\gamma=4$ measurements.
The first line shows the results of the standard GS, the second shows the results of GS2-1, and the third is our method.
The last row is the ground truth image sent.
We used a diverging color map \cite{moreland2009diverging} to highlight the presence of wrong negative pixel values (in red).
}
\end{figure}

The non-interferometric setup employed is a standard implementation of the imaging system used to study disordered media.
We presented its sketched schematics in Fig. \ref{fig:setup}.
A single-mode continuous-wave laser source (vertically polarized) is coupled to a spatial light modulator (SLM), which controls the field that enters the disordered device.
The modulated light is injected into the input facet (I) of a multi-mode fiber and undergoes random scattering events during its propagation.
We record the propagated intensity field on the output facet (O) by selecting the horizontal polarization.
The polarization orthogonal to the laser source guarantees that photons were scattered at least once during their propagation.
We decided to divide the entire SLM area impinging on the input edge in $24\times 24$ segments, and we analyze a squared window of $256 \times 256$ pixels within the central core of the output edge of the MM-fiber.
The region that we use is delimited with a squared box on the speckle pattern of Fig. \ref{fig:schemeGS}.
For the characterization, we send a total of $M_{tot}=10^4$ patterns and, consequently, we record the same number of speckle realizations $S$.
Since we control $N=576$ segments in SLM, it is useful to define the sampling ratio as $\gamma=M/N$.
In this representation, $\gamma=1$ means to sample a number of measurements equal exactly to the number of pixels controlled at the SLM.
These measurements will be used to characterize the TM.
For the imaging procedure, instead, we send $400$ letters randomly extracted from the Greek alphabet.
Few symbols are shown in the bottom row of Fig. \ref{fig:symbolsRec}; these are the objects that we aim at reconstructing.

\section{\label{sec:Results} Results}
After acquiring the measurements, we use the double phase retrieval method to reconstruct the TM and the symbols.
Here, we discuss the results obtained with our protocol against the reconstructions provided by standard PRs described in the literature.
We organize our study as the following.
To retrieve the TM, we made use of binary random patterns only and we run independent phase retrievals using different sampling ratios, %
$\gamma$, ranging from $0.5$ to $10$ in steps of $0.5$.
For the imaging procedure, instead, we use only the speckle patterns obtained by the propagation of the Greek letters through the MM-fiber whose number is $M/25$.
The actual symbols are used to evaluate the quality of the reconstructions after a varying number of iterations.
We use the normalized root-mean-square error (NRMSE) as the metric to compare the reconstructed images with the ground truth.
If we call $p'\left(x,y\right)$ the reconstructed image and $p\left(x,y\right)$ the original object transmitted, we have that:
\begin{equation}
    \text{NRMSE} = 1 - \frac{\text{max} \left| p'\left(x,y\right) \star p\left(x,y\right) \right|^2}{\sum_{x,y} {\left| p'\left(x,y\right) \right|^2} \sum_{x,y} {\left| p\left(x,y\right) \right|^2}}.
\end{equation}
By definition, NRMSE ranges from $0$ to $1$, and a lower value corresponds to a better reconstruction.

\begin{figure}[b!]
\includegraphics[width=1.\linewidth]{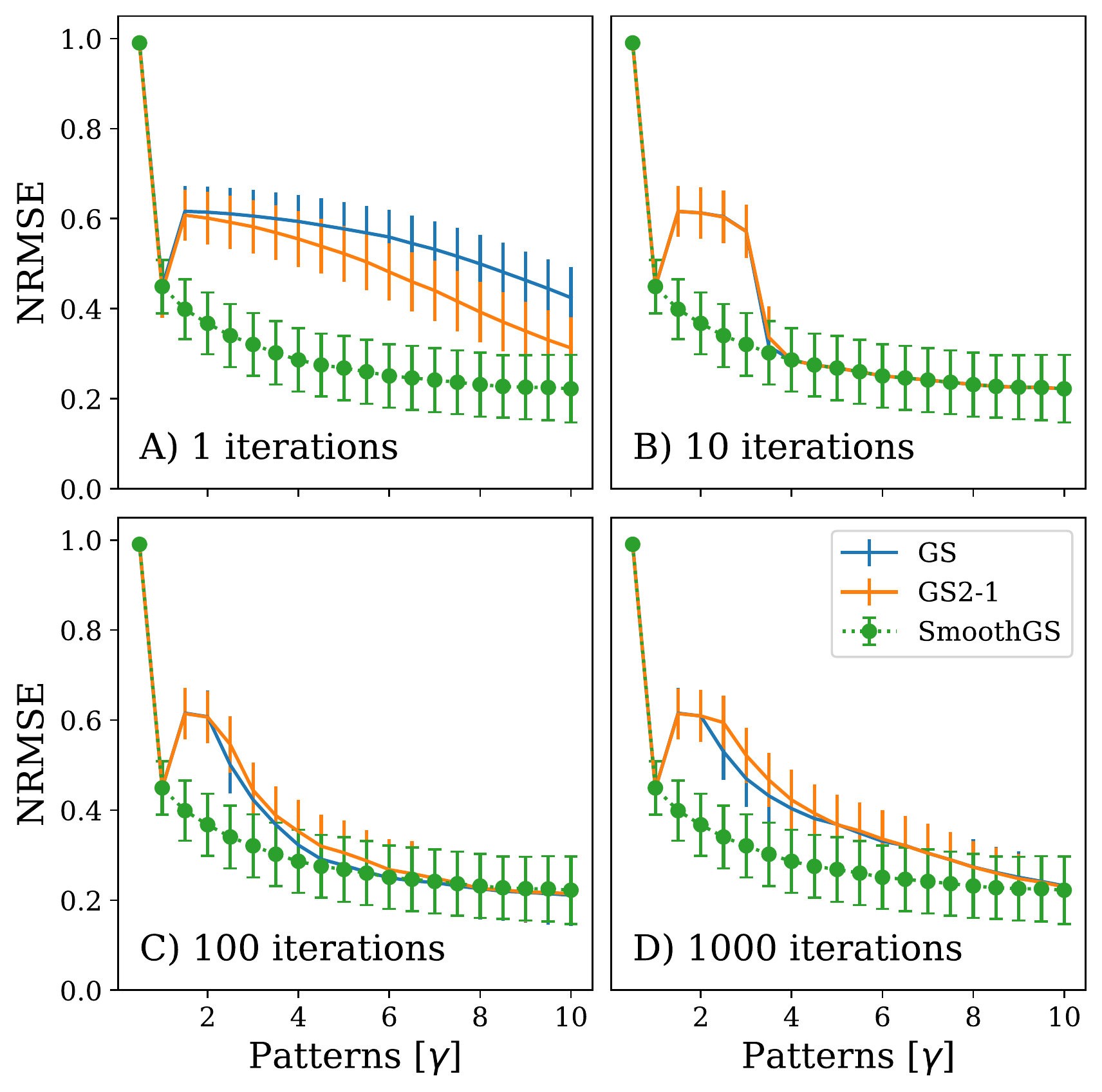}
\caption{\label{fig:performancePlot}
Imaging performance comparison of GS algorithms.
A) results after a single iteration, B) after 10 iterations, C) after 100 iterations, and D) after 1000.
The error bar represents the standard deviation of the image reconstruction over all the objects to be reconstructed.
}
\end{figure}

We begin our study by running independent double-PRs used for reconstructing the test images, varying the number of calibration patterns in combination with a variable number of iterations.
For comparison, we make use of the standard GS and improved GS2-1, proposed by \cite{huang2021generalizing}, that we use as the reference against our method.
\begin{figure*}[t!]
\includegraphics[width=0.95\linewidth]{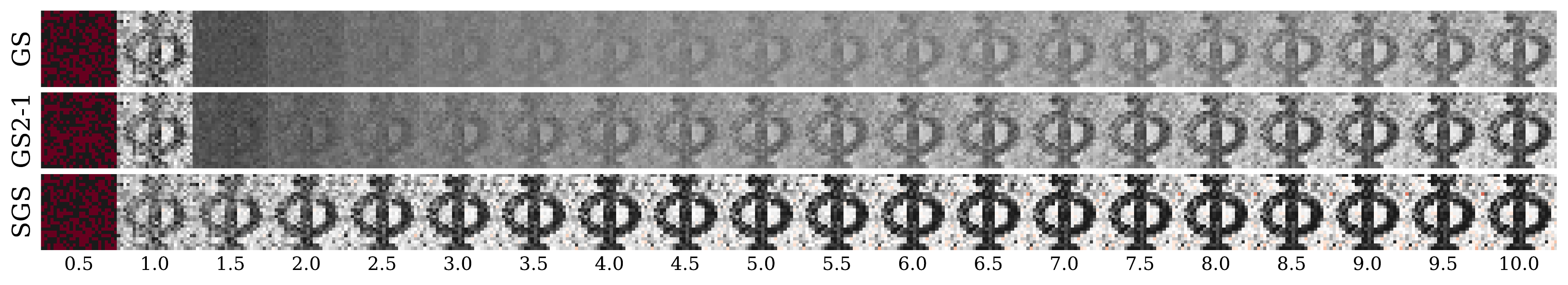}
\includegraphics[width=0.95\linewidth]{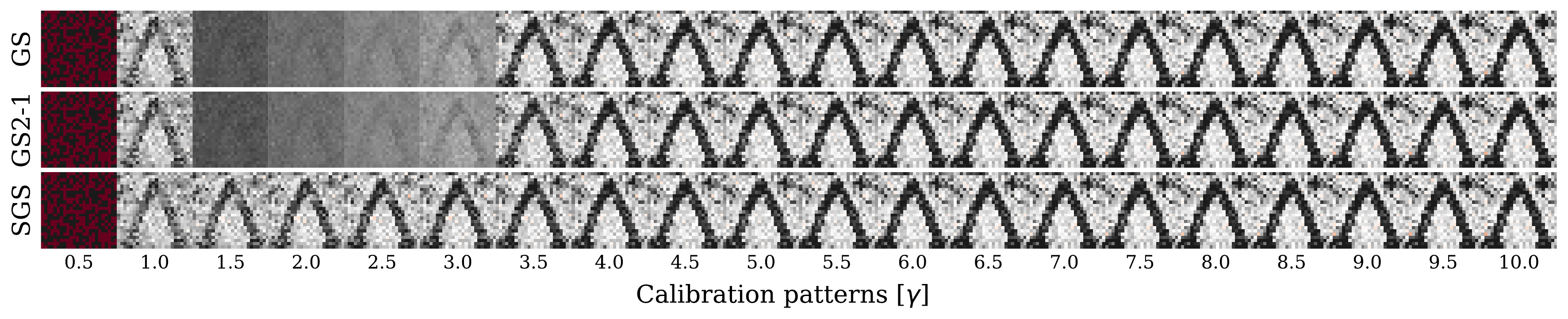}
\caption{\label{fig:imagingAnalysis} Comparative study on the reconstruction of images at progressively increasing sampling ratios.
The block on the top shows the reconstructions after a single iteration.
The bottom block shows the same study after $10$ iterations.
For each block, the first row shows the reconstruction results at various sampling ratios $\gamma$ for the GS.
The second row shows the results of GS2-1 and the third for our method.
}
\end{figure*}
In Fig. \ref{fig:symbolsRec}, we report the image reconstructions of a few symbols recovered after a single iteration of the three GS methods.
In this case, a \textit{single iteration} implies a single GS-step for either the TM reconstruction and the imaging procedure, with a sampling ratio of $\gamma=4$.
The two state-of-the-art GS methods return solutions that are not yet formed, with poor contrast and a noisy background (Fig. \ref{fig:symbolsRec} rows 1 and 2).
Our method (Fig. \ref{fig:symbolsRec}, third row), instead, immediately achieves results much closer to the ground truth solution (Fig. \ref{fig:symbolsRec}, fourth row).
To enrich our analysis, we performed a study by varying the number of the iterations over three orders of magnitude: $1$, $10$, $100$, and $1000$ of steps.
In Fig. \ref{fig:performancePlot}, we plot the different NRMSE obtained at the end of each reconstruction run.
The error bars are calculated over all the symbols considered in each reconstruction experiment.
First of all, for $\gamma<1$, we notice that we never obtain any good results.
For $\gamma=1$, instead, the optimal result is readily provided by all the methods.
Reconstructions at this regime exhibit a constant behavior for any number of iterations.
We point out that, only in this particular case, the sampling matrix $\mathbf{P}$ is squared, and so, the inverse is well defined $\mathbf{P}^\dagger = \mathbf{P}^{-1}$.
Ideally, by increasing the samples used for the reconstruction, one would expect the results to get progressively better.
Instead, the GS and GS2-1 exhibit a performance drop that does not improve by running longer iterations for any $\gamma \in \left(1, 4 \right)$.
In this regime, we note that GS2-1 performs better than standard GS as reported by Huang et al. \cite{huang2021generalizing}.
For $\gamma \geq 4$ and after $10$ iterations, the reconstruction quality gets better, though it progressively deteriorates when increasing the number of iterations.

With our method, instead, we obtain regular performances regardless of the number of iterations.
For any $\gamma$ in Fig. \ref{fig:performancePlot}A, it appears evident that the smoothed implementation of the GS achieves its optimal imaging performance already after a single iteration.
The result is maintained up to thousands of cycles, showing excellent numerical stability.
Remarkably, our method outperforms always the state-of-the-art in the down-sampled regime $\gamma \in \left(1, 4 \right)$.
To ease the analysis, we report the direct imaging performance after $1$ (top group) and $10$ (bottom group) iterations in Fig. \ref{fig:imagingAnalysis}.
After a single iteration, SmoothGS turns to be always better than the other GSs, progressively increasing the reconstruction quality for higher sampling ratios.
After $10$ iterations, the other methods approach the same imaging quality of our implementation, exhibiting a visible discontinuity at $\gamma=3.5$.
\begin{figure}[b!]
\includegraphics[width=1.\linewidth]{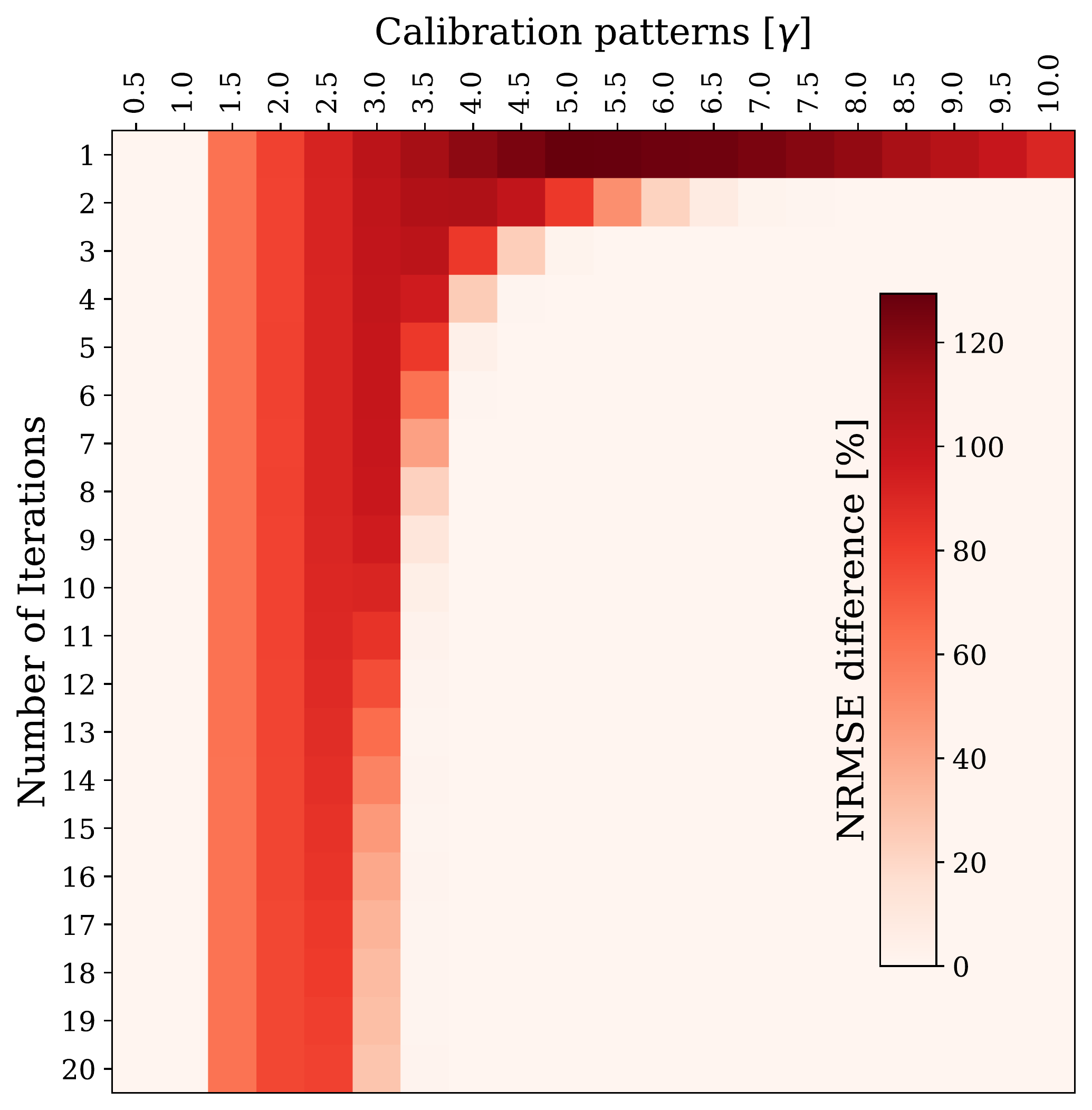}%
\caption{\label{fig:differenceMap} Difference map for the NRMSE of GS and SmoothGS.
Here, we considered a variable number of iterations $\in \left[1,20\right]$.
In the red region, our method always surpasses the current phase retrieval reconstruction.
The color fading to white indicates that the GS method converged to the look-alike reconstruction provided with the single iteration SmoothGS.
}
\end{figure}

From the previous analysis of the plots in Fig. \ref{fig:imagingAnalysis}, we notice that our method is robust and does not improve by running more iterations.
We can consider this to be the optimal result achievable in imaging with GS.
Thus, we use our method as a reference, and we compare the imaging NRMSE obtained with GS2-1 by solving image reconstructions over $1$ to $20$ iterations.
The difference map between SmoothGS and GS2-1 is reported in Fig. \ref{fig:differenceMap}: the whiter the region, the closer the results are.
Again, after a single iteration, our method is unbeaten for any sampling ratio considered.
After two iterations, both methods provide identical results only if $\gamma>7$; after three iterations, the sampling can be decreased to $\gamma>5$.
However, this trend rapidly saturates and, for $\gamma \in \left(1,3 \right)$, no solutions provided by the state-of-the-art methods can approach the image quality provided by ours.

\section{\label{sec:Conclusions} Conclusions}
\begin{figure}[t!]
\includegraphics[width=1.\linewidth]{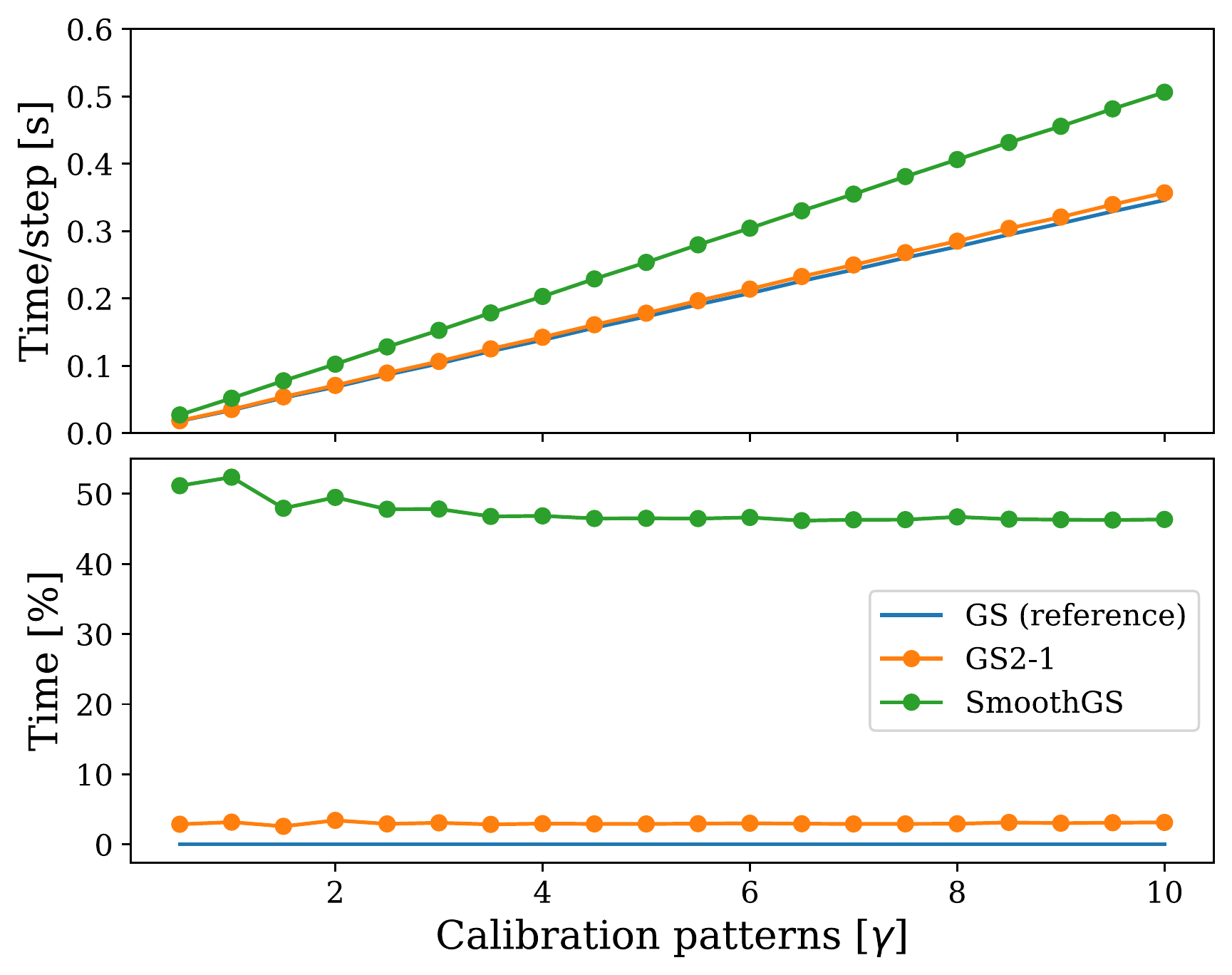}
\caption{\label{fig:temporalExe} Temporal performance of the Gerchberg-Saxton phase retrievals.
In the upper plot, average seconds per step.
In the bottom plot, computational effort variation in percent with respect to the simplest GS implementation.}
\end{figure}

In our article, we described how the introduction of a convolution step in a phase retrieval GS implementation considerably improves reconstruction results.
Although a single iteration of our SmoothGS algorithm is longer than the single iteration in  state-of-the-art phase retrieval algorithms, one is enough to always get a better reconstruction.
Indeed, a single cycle of  SmoothGS consists of 5 steps, whereas the GS and GS2-1 discussed above consist of 4 steps.
For a fair comparison we  test the average computing time of the three methods.
We average $10^3$ iterations comprised of the memory transfer between the CPU and GPU memory.
We report the average seconds per step in the upper plot of Fig. \ref{fig:temporalExe}, and the time variation against the simplest GS in the  plot below.
As expected, GS and GS2-1 almost require the same time to carry out a single iteration, being the second about $3\%$ slower.
Our method, instead, is $50\%$ slower on average on a single  cycle, requiring the convolution step to be carried out.
Even if slower, though, a single SmoothGS iteration is still less expensive than performing two GS iterations (and it might take tens of them for an effective reconstruction, see Fig. \ref{fig:imagingAnalysis}). This grants that our method is always faster in converging than any competitor.
SmoothGS, then, readily provides the best reconstruction achievable with the minimum temporal requirement.

By smoothing the output of the GS, we introduced a model-based regularizer controlled by the statistics of the speckles \cite{goodman2007speckle}.
Our idea comes from noticing that current methods in TM reconstruction neglect neighboring interactions.
The independence of the output pixels permits the separation of the problem so that it requires less memory, and the computational workflow is easily distributable.
However, high memory GPU solutions currently provide enough computational power to carry out the whole task in a single graphic card.
Since our method computes neighboring interactions at the output plane, the whole dataset needs to be processed simultaneously.
Fitting everything in a single GPU avoids the bottleneck of CPU-GPU memory transfer, which would render our method not feasible in terms of computation time.
In particular, the constant development of convolutional neural networks makes it easy to design numerical algorithms which implement depth-wise convolutions in image processing \cite{ancora2021spinning} with PyTorch \cite{paszke2017automatic}.
This constant increase in computational performance is beneficial for our implementation, permitting us to process images with a high number of pixels.

Since TM can be used to engineer the focusing capability of the system \cite{boniface2017transmission}, we can ease its reconstruction when strong non-local correlations are present.
A relevant case is with amorphous speckles \cite{di2015amorphous}, which exhibit quasi-Bessel focusing \cite{di2016tailoring} that implies non-Gaussian kernels.
In these cases, however, a Gaussian fit is not a good choice anymore to recover the smoothing kernel: one may need to invert the speckle auto-correlation to find a good kernel candidate.
Last but not least, the ability to obtain meaningful reconstructions at a low sampling ratio, $\gamma<4$, is a interesting perspective provided by our method.
In fact, it is well known that the TM changes by bending the fiber or varying its temperature \cite{flaes2018robustness}.
In this context, the ability to constantly correct the TM comes in hand with the possibility of reconstruction by using the fewest measurements possible.
In this context, a stochastic choice of the training patterns may help GS algorithms to converge faster \cite{mignacco2021stochasticity}.
The assumption on sparsity in the TM \cite{li2021compressively} and iterative focusing via binary phase-only patterns \cite{geng2021high} have already shown a promising reduction in the number of measurements needed.
The development of fast TM reconstruction frameworks is, indeed, of fundamental interest in biomedical imaging and PR algorithm are promising resources \cite{dong2019spectral}.
With this work, we try to pave the road towards new computational methods that are fast and measurement efficient.
In this respect, our contribution may lead to the definition of new clinical tools for non-invasive and real-time optical measurement.

\subsection*{Competing Interests}
The authors declare no competing interests.

\subsection*{Data Availability}
The data is available upon reasonable request to the authors.

\subsection*{Author Contributions}
D.A. and L.L. conceived the idea and wrote the manuscript.
D.A. designed the code and performed numerical studies.
L.D., A.G., P.C., M.D.G., and D.B. set up the experimental acquisition.
D.S. coordinated the experiments.
L.L. supervised the project.
All the authors reviewed the manuscript.

\begin{acknowledgments}
We acknowledge the support from the European Research Council (ERC) under the European Union’s Horizon 2020 Research and Innovation Program, Project LoTGlasSy (Grant Agreement No. 694925) and Prof. Giorgio Parisi.
We also acknowledge the support of LazioInnova - Regione Lazio under the program {\em Gruppi di ricerca 2020} - POR FESR Lazio 2014-2020, Project NanoProbe (Application code A0375-2020- 36761).
\end{acknowledgments}

\appendix

\section{Description of the experimental setup}
We used a Spatial Light Modulator (SLM, Hamamatsu) to shape the wavefront of a vertically polarized He-Ne continuous-wave laser.
The laser beam was expanded using a two lenses system to a spot size with an approximated radius of $0.4 cm$ and shined onto a portion of the SLM display.
The first-order diffraction from the SLM was collected by a lens and spatially filtered from the other orders, then focused onto the input facet of the MM optical fiber.
A camera image the output edge of the MM-fiber via a two-lens system and a polarizer.
The MM-fiber was thermally isolated in a pipe filled with thermal foam.
The wavefront shaping by the SLM pattern was achieved by superimposing a grating pattern to the binary phase input array.
In this way, the first-order diffraction is shaped into a squared matrix, and we can turn on and off the light at every grid position.
Then, the pattern is sent to the fiber input.
The camera at the entrance allows setting the position of the incoming pattern.
The exit speckle pattern is collected at the other facet by the other camera and saved.
An automatized routine allows synchronizing the proper timing between the two steps.

\section{Image processing of camera pictures.}
The camera acquisition of the facet of the fiber is circular due to its aperture.
From this image, we crop a squared region of $256 px$ inscribed within the facet.
Subsequently, the images were preprocessed by simple intensity normalization.
No further processing was applied to the acquired images.
On the other hand, the fiber input was generated as a random binary pattern.

\section{Convolution and Cross-correlation}
The discrete convolution that we use in the text is defined as:
\begin{equation}
    f * g = \sum_{x,y} f\left(x,y \right) g\left(i-x,j-y \right).
\end{equation}
The definition of the cross-correlation is similar, but $g$ has inverted coordinates:
\begin{equation}
    f \star g = \sum_{x,y} f\left(x,y \right) g\left(i+x,j+y \right).
\end{equation}
By using the previous formula, the autocorrelation can be defined as the cross-correlation of a function with itself, $\chi = f \star f$.

\section{Speckle statistics and their auto-correlation.}
\label{app:C}
Our work is based on the assumption of neighboring coupling of the pixels in the output image.
To estimate the extent of this interaction, we calculate the average auto-correlation of the speckle pattern recorded by the camera.
Since the intensity distribution in the output edge of the MM-fiber could be not uniform, we normalize the intensity speckles dividing them by their slowly varying envelope, as in \cite{katz2014non}.
We estimate this envelope by blurring each camera detection with a Gaussian kernel bigger than the average speckle size.
In this case, we used a standard deviation of $\sigma' = 25px$.
The patterns normalized in this way are auto-correlated, and their auto-correlations are averaged through all the measurements.
In this way, we determine the auto-correlation of the average speckle size.

Successively, we fit the resulting average auto-correlation with a bi-dimensional Gaussian profile, $g\left( \Sigma \right)$.
We make this approximation because the auto-correlation of a Gaussian function, with a given $\sigma$, is another Gaussian with twice the previous standard deviation, $2\sigma$.
In practice, if we take a Gaussian function $g\left( \sigma \right)$ and we compute its auto-correlation, we obtain $g\left( \sigma \right) \star g\left( \sigma \right) = g\left( 2\sigma \right) = g\left( \Sigma \right)$.
Because of this, our method assumes that the PSF starts from half the value of the standard deviation fitted.

\nocite{*}

\bibliography{apssamp}

\end{document}